# Is Twitter a Public Sphere for Online Conflicts? A Cross-Ideological and Cross-Hierarchical Look

Zhe Liu[1],

College of Information Sciences and Technology,
The Pennsylvania State University, University Park, Pennsylvania 16802
zul112@ist.psu.edu

Ingmar Weber
Qatar Computing Research Institute,
PO Box 5825, Doha, Qatar
iweber@qf.org.qa

**Abstract**

The rise in popularity of Twitter has led to a debate on its impact on public opinions. The optimists foresee an increase in online participation and democratization due to social media's personal and interactive nature. Cyber-pessimists, on the other hand, explain how social media can lead to selective exposure and can be used as a disguise for those in power to disseminate biased information. To investigate this debate empirically, we evaluate Twitter as a public sphere using four metrics: equality, diversity, reciprocity and quality. Using these measurements, we analyze the communication patterns between individuals of different hierarchical levels and ideologies. We do this within the context of three diverse conflicts: Israel-Palestine, US Democrats-Republicans, and FC Barcelona-Real Madrid. In all cases, we collect data around a central pair of Twitter accounts representing the two main parties. Our results show in a quantitative manner that Twitter is not an ideal public sphere for democratic conversations and that hierarchical effects are part of the reason why it is not.

## 1 Introduction

With the rapid growth of Twitter, it has become one of the most widely adopted platforms for online communication. Besides using it for relationship formation and maintenance, many people also regularly engage in discussions about controversial issues [1]. On one hand, this increasing adoption of Twitter for online deliberation inevitably creates a perfect environment for open and unrestricted conversations. On the other hand, individuals on Twitter tend to associate more with like-minded others and to receive information selectively. This leads the cyber-pessimist to emphasize the vital role of opinion leaders in shaping others' perceptions during a conflict and to foresee the online environment as a disguise for those in higher social hierarchy to disseminate information. In order to empirically understand whether Twitter creates a public sphere for democratic debates we ask questions like: How do people on different sides of ideological trenches engage with each other on Twitter? How much does social stratification matter in this process? And how universal are such patterns across different types of polarized conflicts?

For our study, we choose three conflicts of very different nature: the Palestine-Israel conflict, the Democrat-Republication political polarization, and the FC Barcelona-Real Madrid football rivalry. Our analysis is guided by four assessment metrics for the democratic public sphere introduced by [2], namely, (i) equality, (ii) diversity, (iii) reciprocity, and (iv) quality. We find that in general Twitter is not an idealized space for democratic, rational cross-ideological debate, as individuals from the bottom social hierarchy not only interact less but also provide lower quality comments in inter-ideological communication. We believe our results advance the understanding of opportunities and limitations provided by Twitter in online conflicts. It is also of relevance for the design and development of conflict intervention tools or procedures as we paint a detailed picture of cross-ideological communication.

[1] Most of this work was done while the first author was at Qatar Computing Research Institute.

## 2 Related Work

The notion of public sphere is defined by Habermas as democratic space for open and transparent communication among publics [3]. In his view, a public sphere was conceived as a space in which: first, communicators are supposed to disregard their social status, so that better argument could win out over social hierarchy. Second, debates should focus on issues of common concerns and should discursively formulate core values. Third, everyone should be able to access and take part in the public debates.

With the advent of the Internet, some optimistic researchers viewed it as a better public sphere than traditional media considering its high reach [4, 5], anonymity [6], diversity and interactivity [2]. In contrast, pessimistic scholars claimed that online discourse oftentimes ends in miscommunication and cannot directly enhance democracy [7]. Also, individuals within the same deliberating group online usually end up at a more extreme position in the same general direction [8, 9] due to selective exposure [10, 11]. In addition, [8] rejected the claim that social stratification is leveled out by the "blindness" of cyberspace, and argued that even in online environment social hierarchy hindered the democratic process of inter-personal communication.

In recent years, the center of the debate has been changed from "Internet as public sphere" to "SNS as public sphere". Optimists argued that the features and tools provided by SNS facilitate communication between individuals, and may be a better means of achieving a true public sphere than anything that has come before it [12, 13]. In contrast, [14, 15] claimed that certain Facebook designs make it a difficult platform for public discourse. In addition, [16-19] noticed that individuals on SNS formed dense clusters that were ideologically homogeneous, although [20] proposed a completely different view, stating that Twitter users tend to share news without bias.

To have a more comprehensive understanding of the afore-mentioned works, in Table 1 we performed a classification of the existing literatures according to the type of platform being studied, as well as Habermas's criteria of public sphere. We colored the literatures to indicate whether it is in support of or against a public sphere.

| Platform | Equality | | Inclusiveness | | Argument Rationality |
|---|---|---|---|---|---|
| | Disregard Hierarchies | Equal Accessibility | Interaction Scale | Interaction Diversity | |
| Website, Blog, Forum | [8] | [4] [6] [8] [21] [2] | [5] [21] [2] | [6] [2] [10] [11] | [21] [2] |
| SNS | [14] | [12] [13] [14] | [12] [13] | [20] [28] [18] [17] [16] [14] [15] | [15] [14] |

**Table 1.** Review of literatures on public sphere. Red indicates evidence in support of a public sphere, blue indicates evidence against.

From that table we saw that: first, most of the existing works mainly focused on online selective exposure, which is just a subcomponent of a healthy public sphere according to Habermas's conception. Second, although comprehensive assessments have been conducted on blogs and forums as public spheres [2, 8, 21], we argue that one cannot simply map these findings onto SNS, due to its very different network structures and communication features. Last but not least, social hierarchy, as a very important criterion in evaluating public sphere, has rarely been addressed in prior literatures. Thus, in this study we want to determine among others, if social hierarchy has an effect on individual's participation in democratic communication on Twitter.

## 3 Research Questions

In this work, we aim to assess if Twitter is a public sphere for democratic debates. We used Habermas's conception as the theoretical framework for our analysis and evaluate each of those dimensions with the assessment metrics: equality, diversity, reciprocity and quality, proposed by Schneider [2]. Figure 1 depicts the research framework of this study.

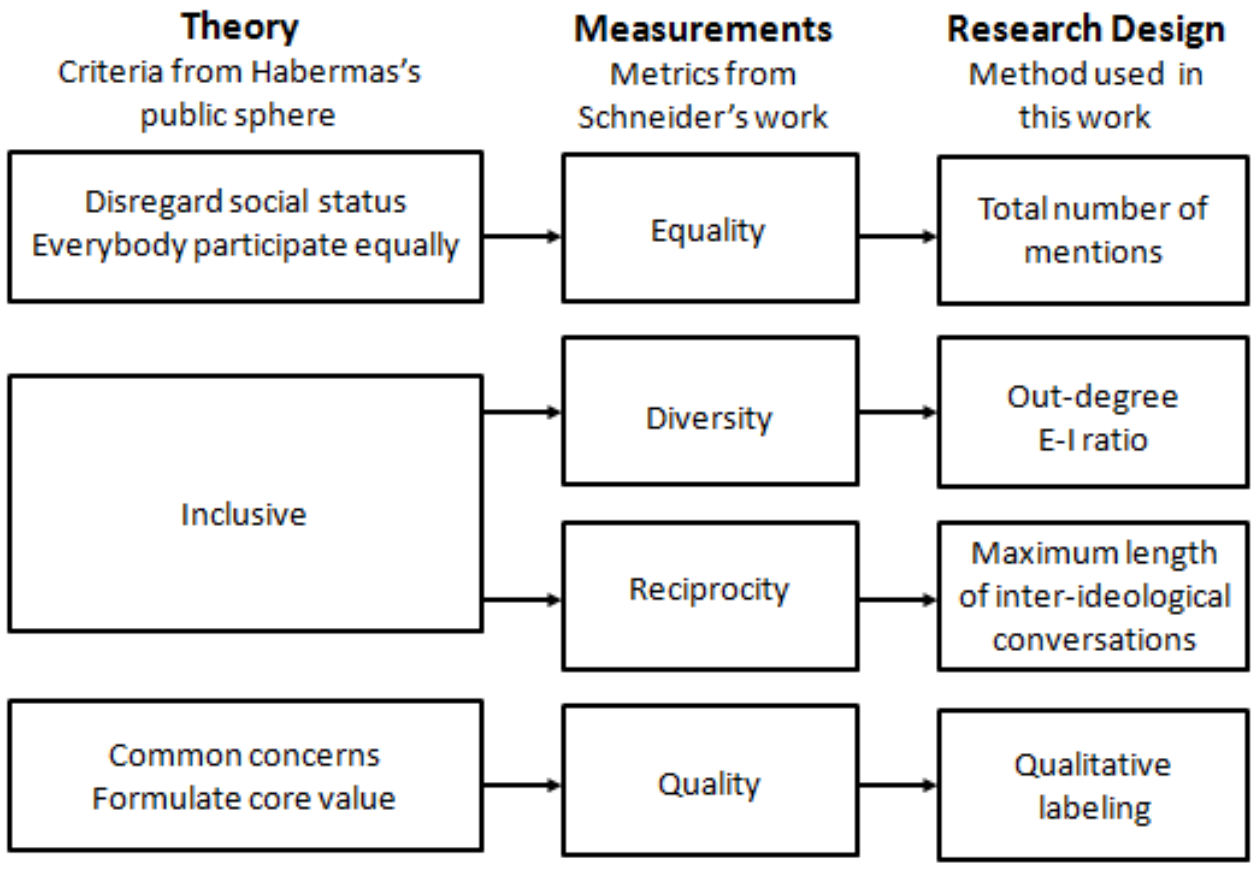

**Fig. 1.** Research framework for evaluating Twitter as a public sphere.

- ***Equality.*** A democratic state requires all individuals, regardless of their social status, to engage and contribute equally in communication [2]. We quantify a user's engagement by the total number of mentions they make to ideological-friends or foes.
- ***Diversity.*** A healthy public sphere requires a diverse communication network, which suggests the flexibility of an individual in adapting to varied opinions and views [30]. We adopt the measurement called external-internal (E-I) index to measure the diversity of one's communication. The E-I index is calculated as:

$$\mathrm{EI_i} = \frac{\mathrm{E_i} - \mathrm{I_i}}{\mathrm{E_i} + \mathrm{I_i}}$$

where Ei is number of unique ideological-foes user i has interacted with, Ii is the number of unique ideological-friends user i has interacted with. The E-I index ranges from -1 to 1. The closer the E-I index is to -1, the more an individual tends to only talk to members of their own group, suggesting a high degree of insularity.

- ***Reciprocity.*** High reciprocal interactions promote the dyadic exchange of information and resources among individuals, and thus ensure a democratic communication environment. To evaluate the reciprocity levels across ideologies and hierarchies, we adopted the maximum length of inter-ideological conversations as the measurement. We chose the maximum over average in order to avoid the bias introduced due to Twitter API's restriction of getting more than 3,200 tweets per user.
- ***Quality.*** High-quality communication requires participants to be polite to each another, even during disagreements. Besides, it also encourages participants to make rational arguments supported by logical explanations. High-quality political discourse is important to building democratic consensus. Quality is measured using a crowd-sourcing method, which we will discuss in more details in later sections

## 4 Method

### 4.1 Data Collection and Labeling

To automatically detect users with similar or different ideologies, we started with three pairs of opposing seed users: *@AlqassamBrigade* and *@IDFSpokesperson*, *@TheDemocrats* and *@GOP*, and *@FCBarcelona_es* and *@realmadrid*. We intentionally chose these accounts as seed nodes due to their key roles in well-known real-life conflicts which are also reflected on Twitter. For each of the seed nodes, we obtained up to 3,200 of its latest tweets using the Twitter API. For each tweet, we identified up to 100 of its retweeters and labeled them as likely supporters. We use retweet as a signal for ideological categorization by following [22], as retweet usually represents one's endorsements and preferences [23]. We removed mediators and neutral intervenors, such as peace movement organizations and journalists, from our datasets based on their distinct retweeting patterns by following the method introduced in [24].

Classification results were validated via CrowdFlower [25] by assigning 100 random users in each ideology to the HIT workers. By comparing user's pre-assigned ideology to the majority-voted label obtained from CrowdFlower, we found that our classification method yielded on average an accuracy of 96.2%. With the classified users, we extracted all mentions between them as interactions between ideological-friends and foes. Table 2 lists the descriptions of our collected datasets. In total we collected 226,239 Twitter users involved in all three conflicts. Among over 400 million of their daily tweets, we extracted only tweets containing cross- or within-ideological interactions from 56,024 unique users. While comparing the inter- and intra-party tweets, we noticed that they are far less interactions between ideological-foes than friends.

| Conflict | # Users | #Intra-Mentions | #Inter-Mentions | #Intra-Retweets | #Inter-Retweets |
|---|---|---|---|---|---|
| **PA – IL** | 9,937 | 42,471 | 3,772 | 135,784 | 1,057 |
| **DEM – REP** | 17,869 | 105,557 | 16,927 | 471,291 | 3,330 |
| **DEM – REP** | 17,869 | 105,557 | 16,927 | 471,291 | 3,330 |
| **FCB – RMCF** | 28,218 | 47,924 | 7,996 | 104,875 | 13,093 |

**Table 2.** Datasets Statistics

In addition to dividing the collected users into two camps for each dataset, we also split them into four social hierarchical groups according to their number of followers, including: the top 1%, the 1% - 10%, 10% - 70%, and 70% - 100% users. The division is arbitrary, but we think that the number of followers at least partly indicates a person's degree of influence on the social network [26], even though it may not fully represent the social status of an individual in real world.

### 4.2 Analysis of Inter and Intra-ideological Communication

To test our first three hypotheses, regarding the equality, diversity and reciprocity across hierarchical levels, one-way ANOVA tests were conducted, with significance level set at 0.05. Post-hoc analyses were also carried out with Tamhane's T2 test due to non-homogenous variances. Prior to analysis, all data were checked for normality and non-normal data was transformed using the Log(x+1) method.

| **Openness of Attitude** | |
|---|---|
| **Agree** | A tweet that agrees with the other user or shows similar opinions on the covered material. |
| **Neutral** | A tweet that is neutral in nature, neither in obvious agreement or disagreement. |
| **Disagree** | A tweet that disagrees with (or critiques) the other user (or the party him/her supports) or shows different opinions on the covered material. |
| **Insult or Sarcasm** | A tweet that can be regarded as a derogatory message, such as curses, insults, personal abuse, sarcasm or words that indicated pejorative speak. |
| **Off-Topic** | A tweet that is totally unrelated to the conflict. |
| **Unclear** | A tweet that is does not fall into any of the above categories. |
| **Rationality of Argument** | |
| **Highly rational** | The user used information from external sources and with statements based on facts or data, etc. |
| **Rational** | The user claimed based on his/her viewpoint and with fair and logical argument to support the statement. |
| **Irrational** | The user claimed based on subjective arguments with-out any kind of validation or presentation of facts. |

**Table 3.** Coding scheme for communicator's attitude and rationality.

For the hypothesis of communication quality, we again relied on CrowdFlower. We analyzed the quality of inter-ideological conversations from two perspectives, including the openness of the communicator's attitude, and the rationality of his/her argumentation. To be more specific, for each combination of the four hierarchical levels, we randomly sampled 50 user pairs with cross-ideological conversations. Next, for each of the user pairs, we extracted one of their complete conversations and highlighted one tweet in it at random. We displayed the selected conversation to the workers. From reading the highlighted tweet, we asked them to label the user's attitude and rationality according to our pre-defined coding schemes as shown in Table 3. To provide more contexts, the user's profiles as well as their automatically detected ideologies are also displayed in the HIT.

## 5 Results

### 5.1 Descriptive Results of Cross-Hierarchical Communication

To examine the social hierarchical effect on cross and within-ideological communication patterns, we calculated the conditional probability of a communicator interacting with another, given their social hierarchies. Note that due to the conditional probabilities, different activity levels of the different tiers do not affect our results. Here we used the PA–IL conflict for illustration purpose and only reported findings that can be generalized to all three datasets. As shown in Figure 1, the horizontal bars depicted the four social hierarchical levels of the conversation starter / receiver. The width of the bar denoted the number of interactions existed within that level.

From the width of the horizontal bars in Figure 2, we saw that except the bottommost level, users from the other three hierarchies have about the same probabilities of being mentioned by their ideological-friends. However, under an inter-ideological context, we noticed that users in the topmost hierarchical level have the highest chance of receiving a mention initiated by their foes, which is even higher than the sum of the probabilities derived from rest three levels. This indicated that people are more willing to attack or challenge "authorities" in online conflict. Besides, under both conditions, there is very little chance that the bottom users will be mentioned by either their friends or foes. In addition, from viewing the width of all ribbons, we found that users from the bottommost hierarchical level maintain the highest probability of initiating a mention of the top 1% of users.

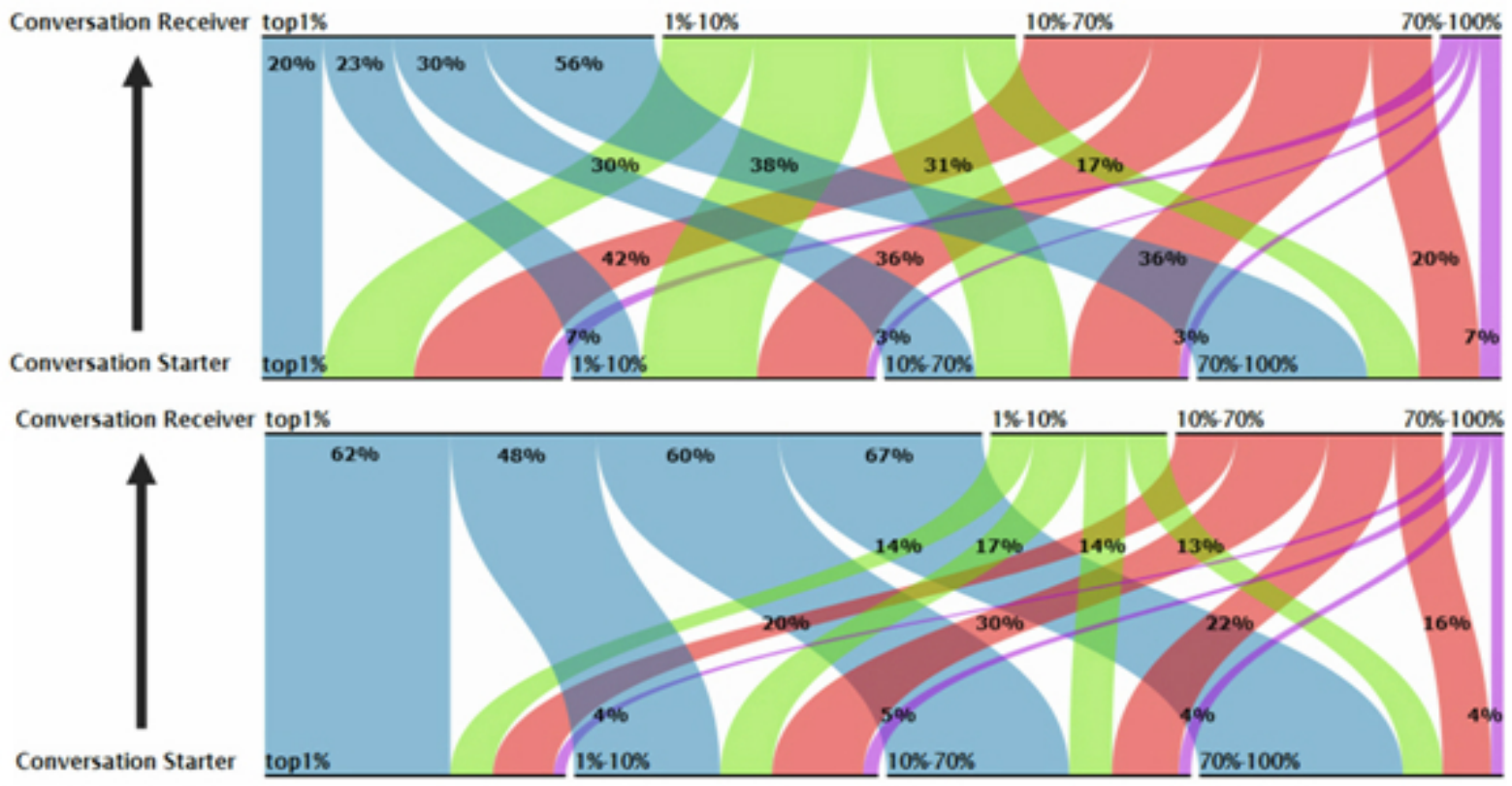

**Fig. 2.** Inter (below) and intra-ideological (above) communication across hierarchies for the PA-IL dataset.

### 5.2 Twitter as Public Sphere

This section presents the findings with regard to each of our proposed measurements. As similar patterns were observed for the two political datasets, only the analysis results from the PA – IL conflict would be shown below for illustration purpose.

**Equality.** For equality measurement, we categorized users into groups as introduced in [27] based on their number of inter- and intra-ideological mentions. We noticed from Table 4 that users in the upper hierarchical levels initiated more conversations with their ideological-friends than those from the lower levels. The ANOVA results further indicated that these differences were significant at the 5% level (PA-IL: $F = 119.12$, $p = 0.00$; DEM-REP: $F = 530.34$, $p = 0.00$). We assumed that this might be relevant to the political celebrities' intentions of maintaining their position and status, as well as to stay connected with their supporters, although this needs to be proved in future studies. FCB-RMCF dataset revealed very different results, with only the bottom users initiated more conversations than users from the upper levels.

When analyzing the inter-ideological communications, we did not find such differences across social hierarchies within the PA-IL ($F = 0.73$, $p = 0.41$) and FCB-RMCF conflict ($F = 0.59$, $p = 0.63$). Although the ANOVA results on the DEM-REP dataset was significant ($F = 27.45$, $p = 0.00$), from the results of the Tamhane's T2 test we further noticed that only users in the bottom group were involved in significantly less interactions with their ideological-foes. In that sense, we claim that at least from our experiments, Twitter allows individuals to disregard their social status in real world, and facilitates their equal participation in online political discourse.

| Participation Type (# of mentions) | #Users with Intra-ideological Mentions | | | | #Users with Inter-ideological Mentions | | | |
|---|---|---|---|---|---|---|---|---|
| | 1% | 1%-10% | 10%-70% | 70-100% | 1% | 1%-10% | 10%-70% | 70-100% |
| One time (1) | 1 (1.1%) | 35 (4.3%) | 435 (9.5%) | 359 (19.6%) | 6 (19.4%) | 32 (20.5%) | 198 (20.9%) | 110 (24.8%) |
| Light (2-5) | 6 (6.8%) | 95 (11.8%) | 1002 (21.8%) | 579 (31.7%) | 8 (25.8%) | 42 (26.9%) | 292 (30.9%) | 147 (33.2%) |
| Medium (6-20) | 16 (18.2%) | 220 (27.2%) | 1307 (28.5%) | 485 (26.5%) | 8 (25.8%) | 37 (23.7%) | 239 (25.3%) | 95 (21.4%) |
| Heavy (21-79) | 31 (35.2%) | 274 (33.9%) | 1117 (24.4%) | 307 (16.8%) | 7 (22.6%) | 30 (19.2%) | 172 (18.2%) | 66 (14.9%) |
| Very Heavy (80+) | 34 (38.6%) | 184 (22.8%) | 726 (15.8%) | 99 (5.4%) | 2 (6.5%) | 15 (9.6%) | 45 (4.8%) | 25 (5.6%) |

**Table 4.** Equality of participation across social hierarchies (PA-IL)

**Diversity.** The one way ANOVA tests on E-I index showed significant differences for all three datasets (PA-IL: F = 25.29, p = 0.00; DEM-REP: F = 24.06, p = 0.00; and FCB-RMCF: F = 62.34, p = 0.00), with the second hierarchical group of both political datasets had the significantly lowest E-I index, indicating that people in that social hierarchy are more insular toward their ideological-foes. In contrast, the bottom hierarchy exhibited the highest tendency towards inter-ideological communications. Unlike the political datasets, our post-hoc analysis on the sports dataset again demonstrated completely different patterns of insularity, with the bottom users more willing to interact within their own camps. Consistent with findings from prior studies [16-19], all E-I index were less than 0, indicating individual's preferences of talking to their ideological-friends.

**Reciprocity.** The ANOVA tests also indicated significant overall differences on the maximum length of intra-ideological conversations across hierarchies (PA-IL: F = 86.32, p = 0.00; DEM-REP: F = 807.15, p = 0.00; FCB-RMCF: F = 355.706, p = 0.00), with the maximum frequency of back-and-forth communications increased along with the level of the conversation starter's social hierarchy. In other words, when talking to friends with higher social status, people tended to show greater reciprocity.

However, when analyzing the reciprocity in cross-party debates, the ANOVA and post hoc tests showed no (DEM-REP: F = 27.56, p = 0.06) or almost no (PA-IL: F = 4.60, p = 0.00; FCB-RMCF: F = 10.24, p = 0.00) significant effect of social hierarchy on conversation reciprocity, with only conversation starters from the bottom hierarchy had significantly less back and forth exchanges in cross-ideological conversations.

**Quality.** Table 5 lists the annotation results on inter-ideological communications. The analysis results of the FCB-RMBC dataset were not included here, as the majority of the inter-ideological conversations within that conflict are off-topic chit-chats. For the two political datasets, we found that "disagreement" tweets dominated all the inter-ideological discussions, accounting for more than 70% of all posts. "Insults or sarcasm" were the second most common communication type identified. About 8% of all arguments were personal attacks. 46.7% of all invective posts were from individuals in the bottom level. Inter-party agreements were fairly rare in our results.

| Conflict | Agree | Insult | Neutral | Off-Topic | Unclear | Disagree | | |
|---|---|---|---|---|---|---|---|---|
| | | | | | | Highly-rational | Rational | Irrational |
| PA-IL | 1.1% | 7.2% | 1.6% | 1.1% | 18.1% | 4.1% | 63.7% | 3.0% |
| DEM-REP | 3.4% | 8.4% | 1.7% | 2.0% | 9.2% | 4.7% | 67.6% | 3.0% |

**Table 5.** Statistics on inter-ideological communication types and rationality.

Next, in our analysis of the argument rationality, we first noticed that the majority of people (89.9%) in inter-ideological discussions demonstrated at least some rational attempts to justify their viewpoints to opponents. Irrational arguments were detected in only 5.8% of all conversations. Highly rational statements were even rarer, accounting for only 4.3% of all annotated tweets. We noticed 31.6% of all statements with highly rational argument were from the top 1% of users.

| | 1% | 1-10% | 10-70% | 70-100% |
|---|---|---|---|---|
| **Urls to Foes** | 19 (63.3%) | 80 (54.1%) | 430 (49.5%) | 151 (22.2%) |
| **Equal Urls** | 3 (10.0%) | 27 (18.2%) | 227 (26.2%) | 158 (42.1%) |
| **Urls to Friends** | 8 (26.7%) | 41 (27.7%) | 211 (24.3%) | 66 (20.2%) |

**Table 6.** Type of rationality across social hierarchies

Assuming that rational individuals tend to rely on external resources to support their viewpoints, we also quantified users' rationality in this section by measuring the differences in the percentage of URL usages between inter- and intra-ideological mentions. Based on such percentage differences, we categorized all individuals into three groups: more URLs shared with ideological-friends, with ideological foes, and equal URLs shared with both ideological friends and foes. As shown in Table 6, we found that more than half of the individuals from the first two social hierarchical groups adopted more URLs when talking to ideological foes, whereas individuals in the bottom social hierarchy tended to be more rational to their ideological friends.

To further explore the differences in content between inter and intra-ideological conversations, we generated a word cloud in Figure 3 with the top 50 words with the largest relative differences in the usage probabilities. The font size in this word cloud correlates with the absolute difference of a word occurring with a higher probability in only one of the two classes. We colored words that appeared more in inter-ideological conversations blue and otherwise red. We found that, first, blue words are in general larger than the red ones, indicating that inter-ideological talks stick more to the controversial topics compared to the intra-ideological ones. Second, it is very clear that words adopted in inter-ideological conversations are more negative (e.g. "kill", "murder", "hate") in tone compared to words in intra-ideological talks (e.g., "thank", "great", "love").

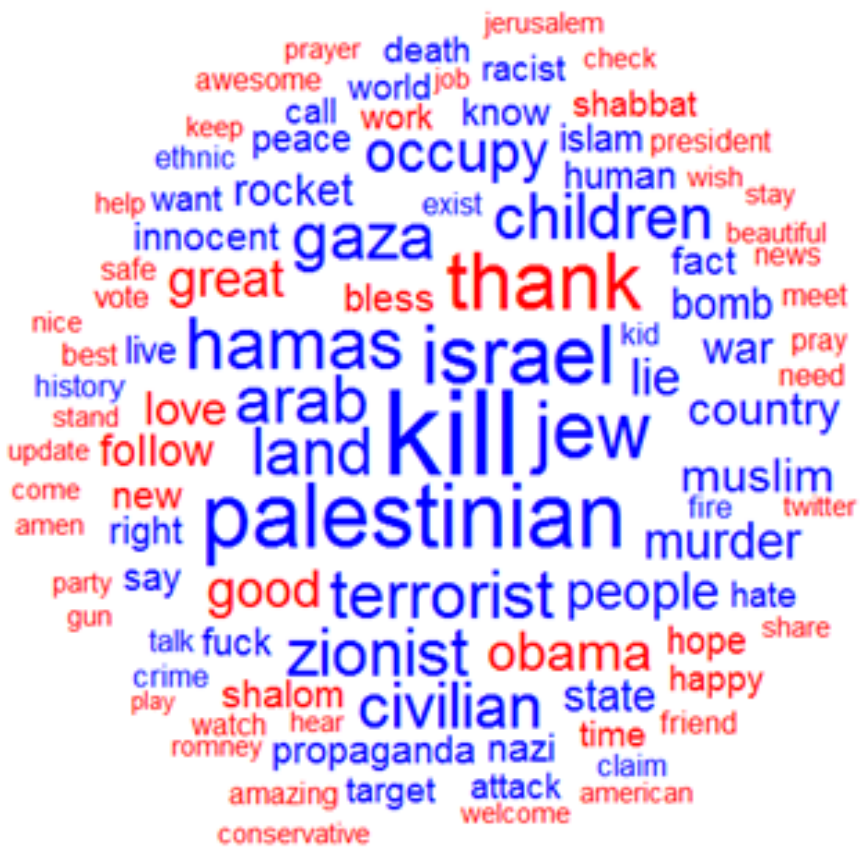

**Fig. 3.** Tag cloud with relative differences in inter- vs. intra-ideological usage probabilities for the PA-IL dataset.

## 6 Discussion and Conclusion

Through our analyses on three datasets, we concluded that individuals demonstrated inconsistent communication behaviors in conflicts of different natures (political vs. sports). Ideological and social status played important roles in shaping one's communication habits in political conflicts, and in the meanwhile posed a challenge for conducting democratic discussions on Twitter. First, our work also found selective exposure a problem in Twitter conflicts, as users are more willing to share and to communicate with their ideological-friends than foes. Second, we noticed that most of the cross-ideological mentions on Twitter were initiated toward political authorities in higher social hierarchies, whereas the general public in the bottom hierarchy were mostly ignored. Third, in general the duration of a within ideological conversation was longer than that of a cross-ideological one. Also, conversations initiated by the top users tended to last longer than those initiated by the bottom ones. Fourth, in our experiment more than 40% of cross-ideological tweets were disagreements. This leads us to think Twitter's failure in facilitating the establishment of cross-party agreements.

Although Twitter cannot be viewed as a public sphere for the above issues, we believe it still has a great potential in becoming a platform for resolving online conflicts. Through our analysis of equality we found that Twitter users disregard their social status, participated equally in cross-ideological communications. Additionally, to our surprise, there are very few insulting tweets labeled in our experiments. Most of the arguments on Twitter are claims based on rational viewpoint, though without referencing any external source, fact or data. We think this kind of logical argumentation can still help spread information or knowledge across-ideologies. As more and more communication happens online and publicly through social media, we deem such an analysis a valuable step towards understanding conflicts online. Understanding the online dynamics of such communication could, among other things, contribute to identifying appropriate mediators to resolve the conflict both online and offline.